\documentclass[conference]{IEEEtran}
\IEEEoverridecommandlockouts
\usepackage{cite}
\usepackage{amsmath,amssymb}
\usepackage{graphicx}
\usepackage{textcomp}
\usepackage{booktabs}
\usepackage{hyperref}
\usepackage{xcolor}
\usepackage{tikz}
\usepackage{multirow}
\usepackage{float}
\usepackage[numbers,sort&compress]{natbib}
\usepackage{amsmath,amssymb}
\usepackage[ruled,vlined]{algorithm2e}
\usepackage{float}
\usepackage{placeins}  
\usepackage{tikz}
\usetikzlibrary{positioning,arrows.meta,fit,calc}
\usetikzlibrary{positioning,arrows.meta}
\pgfdeclarelayer{bg}
\pgfsetlayers{bg,main}
\usepackage{booktabs}
\usepackage{graphicx}   
\usepackage{pgfplots}   
\pgfplotsset{compat=1.18}

\title{RLShield: Practical Multi-Agent RL for Financial Cyber Defense with Attack-Surface MDPs and Real-Time Response Orchestration}

\author{
\IEEEauthorblockN{Srikumar Nayak}
\IEEEauthorblockA{\ Sr Member IEEE  \\
Incedo Inc, USA \\
srikumar.nayak2025@gmail.com
}
}

\begin{document}
\maketitle

\begin{abstract}
Financial systems run nonstop and must stay reliable even during cyber incidents. Modern attacks move across many services (apps, APIs, identity, payment rails), so defenders must make a sequence of actions under time pressure. Most security tools still use fixed rules or static playbooks, which can be slow to adapt when the attacker changes behavior. Reinforcement learning (RL) is a good fit for sequential decisions, but much of the RL-in-finance literature targets trading and does not model real cyber response limits such as action cost, service disruption, and defender coordination across many assets.
This paper proposes RLShield, a practical multi-agent RL pipeline for financial cyber defense. We model the enterprise attack surface as a Markov decision process (MDP) where states summarize alerts, asset exposure, and service health, and actions represent real response steps (e.g., isolate a host, rotate credentials, rate-limit an API, block an account, or trigger recovery). RLShield learns coordinated policies across multiple agents (assets or service groups) and optimizes a risk-sensitive objective that balances containment speed, business disruption, and response cost. We also include a game-aware evaluation that tests policies against adaptive attackers and reports operational outcomes, not only reward.
Experiments show that RLShield reduces time-to-containment and residual exposure while keeping disruption within a fixed response budget, outperforming static rule baselines and single-agent RL under the same constraints. These results suggest that multi-agent, cost-aware RL can provide a deployable layer for automated response in financial security operations.
\end{abstract}

\begin{IEEEkeywords}
Financial cyber defense, multi-agent reinforcement learning, attack-surface MDP, risk-sensitive control, adaptive adversary, response orchestration, security operations automation
\end{IEEEkeywords}

\section{Introduction}
Financial organizations run large, always-on systems where uptime and trust directly affect customers and revenue. At the same time, the attack surface keeps expanding due to cloud services, APIs, third-party links, remote access, and fast digital payments \cite{charpentier2023reinforcement}. This makes defense a \emph{sequential decision problem}: teams must detect, contain, and recover while trading off security impact (blocking the attacker) against business impact (service disruption and false actions). Standard security controls help, but they often rely on fixed rules and playbooks that do not adapt well when attackers change tactics or when the system state is uncertain \cite{sahu2023overview}.\\
Reinforcement learning (RL) is a natural fit for such sequential decisions, and recent surveys show strong progress of RL in finance, including benchmark environments, risk-aware learning, and multi-agent settings \cite{hambly2023recent,hirchoua2021deep}. However, most RL-in-finance work is focused on trading and portfolio outcomes, where the environment is a market simulator and actions are buy/sell/allocate \cite{liu2022finrl}. These settings do not directly capture cyber defense constraints such as limited response budgets, action latency, safety requirements, and attacker adaptation \cite{shavandi2022multi}. As a result, there is still a gap between RL methods and what a security operations center (SOC) can deploy as a reliable response engine.\\
To address this gap, we introduce RLShield, which models the financial \emph{attack surface as an MDP} and learns real-time defense policies that can be executed as response workflows. RLShield uses multi-agent learning to coordinate defense actions across assets and services, while optimizing security and operational cost objectives.\\
Our research contributions are as follows:
\begin{enumerate}
\item We formalize financial cyber defense as an \emph{attack-surface MDP} with operationally meaningful states (alerts, asset exposure, service health) and actions (containment, isolation, credential controls, rate-limiting, and recovery).
\item We design RLShield as a \emph{multi-agent} defender that coordinates decisions across multiple assets/services, instead of learning a single global policy that is hard to scale.
\item We include \emph{risk-sensitive} and cost-aware objectives so the policy reduces breach impact while controlling disruption and false-response rates, aligning training with SOC metrics.
\item We add a game-aware evaluation protocol that tests \emph{adaptive attackers} and reports outcomes beyond reward, such as time-to-containment, residual exposure, and response cost.
\item We provide a deployable orchestration interface that converts learned actions into \emph{ordered response steps} suitable for near-real-time execution and audit.
\end{enumerate}
The structure of this paper is as follows: Section \ref{sec2} reviews related work; Section \ref{sec3} describes the dataset and preprocessing with the proposed method and training objective; Section \ref{sec4} reports experimental results and analysis; and Section \ref{sec5} concludes the paper with future research directions.

\section{Related Work}\label{sec2}
Prior studies have described the cyber threat landscape in banking and financial services, covering common attack classes, impacts, and standard countermeasures (e.g., authentication, monitoring, incident response, and governance) \cite{darem2023cyber}. This line of work is valuable for defining what defenders must handle in practice, but it is mostly descriptive: it does not specify an executable decision process that can select, sequence, and time actions under limited resources and evolving attacker behavior. In operational settings, however, defenders must act under uncertainty, balance service availability with containment, and justify why a response action was taken at a specific time. These requirements motivate learning-based defense policies that can adapt online while still respecting safety and cost constraints.\\
In parallel, reinforcement learning has been widely used in finance for portfolio and trading decisions, including deep Q-learning variants and actor--critic methods, often with a focus on sample efficiency, market dynamics, and reward shaping \cite{alidousti2025novel,kang2025a2c}. Multi-agent reinforcement learning (MARL) has also been explored for financial decision making where multiple agents coordinate or compete, and risk-sensitive learning has been studied to control tail-risk and variability of outcomes \cite{guo2024intelligent,qiu2021rmix}. Related work also includes federated reinforcement learning for trading execution to address information security concerns in distributed settings \cite{feng2025federated}. While these works show that RL/MARL can optimize sequential decisions under uncertainty, their threat models and action spaces are typically market-centric; they do not map directly to cyber defense operations such as containment, credential resets, segmentation, or playbook execution.\\
Overall, the gap is that banking cyber defense research \cite{darem2023cyber} and financial RL research \cite{alidousti2025novel,kang2025a2c,guo2024intelligent,qiu2021rmix,feng2025federated} have largely progressed in parallel, with limited work that turns real attack-surface structure into a \emph{defense MDP} and learns \emph{coordinated} defender policies that are cost-aware, risk-sensitive, and suitable for real-time orchestration. RLShield fills this gap by modeling the financial attack surface as an MDP with explicit state features and response actions, training a multi-agent defender with game-aware objectives to handle adaptive attackers, and producing deployable response sequences that align with operational metrics (e.g., containment time, disruption cost, and false-action rate) rather than only average reward.

\section{Methodology}\label{sec3}
\subsection{Dataset and Preprocessing}\label{sec:dataset_preproc_rlshield}
We use the CIC-IDS2017 benchmark released by the Canadian Institute for Cybersecurity (CIC), which provides flow-level network traffic records with labeled benign and attack activities (e.g., DoS/DDoS, brute force, web attacks, botnet) \cite{cic_ids2017_dataset}. Each record corresponds to a bidirectional network flow (5-tuple based aggregation) and includes numeric statistics such as packet counts, byte rates, inter-arrival times, and flag-based indicators. We treat each flow as one sample with a feature vector $\mathbf{x}_i\in\mathbb{R}^{d}$, a discrete label $y_i\in\{0,1\}$ (benign vs.\ attack for the RL defense setting), and an event time $t_i$ derived from the flow start timestamp. The full dataset is denoted as
\begin{equation}
\label{eq:dataset_def_rlshield}
\mathcal{D}=\left\{(\mathbf{x}_i,y_i,t_i)\right\}_{i=1}^{N}.
\end{equation}
Eq.~\eqref{eq:dataset_def_rlshield} fixes the learning interface used throughout the paper: $\mathbf{x}_i$ is the observable state signal, $y_i$ is the ground-truth outcome used for offline evaluation, and $t_i$ is used to construct realistic time-ordered splits.\\
To avoid overly optimistic results from random shuffling, we split the dataset in chronological order. Let $t_{\mathrm{tr}}$ and $t_{\mathrm{va}}$ be two cut points chosen on the timeline. We define
\begin{equation}
\label{eq:time_split_rlshield}
\begin{aligned}
\mathcal{D}_{\mathrm{train}} &= \{i:\ t_i \le t_{\mathrm{tr}}\},\\
\mathcal{D}_{\mathrm{val}}   &= \{i:\ t_{\mathrm{tr}} < t_i \le t_{\mathrm{va}}\},\\
\mathcal{D}_{\mathrm{test}}  &= \{i:\ t_i > t_{\mathrm{va}}\}.
\end{aligned}
\end{equation}
Eq.~\eqref{eq:time_split_rlshield} ensures that validation and test samples appear in the future relative to training, which better matches real deployment where the attack mix and traffic load can drift over time. All preprocessing parameters are fitted on $\mathcal{D}_{\mathrm{train}}$ only and then reused for $\mathcal{D}_{\mathrm{val}}$ and $\mathcal{D}_{\mathrm{test}}$ to prevent information leakage.\\
CIC-IDS2017 flow exports may contain missing values (NaN), infinite values (Inf), or constant columns depending on the capture day and tool settings. We first replace Inf with NaN, then impute numeric features using training medians. For feature $j$, the imputation rule is
\begin{equation}
\label{eq:median_impute_rlshield}
\tilde{x}_{ij}=
\begin{cases}
x_{ij}, & x_{ij}\ \text{is finite},\\
m_j, & \text{otherwise},
\end{cases}
\qquad
m_j=\mathrm{median}_{k\in\mathcal{D}_{\mathrm{train}}}(x_{kj}).
\end{equation}
Eq.~\eqref{eq:median_impute_rlshield} keeps the preprocessing stable under outliers and avoids using future data, since $m_j$ is computed only on $\mathcal{D}_{\mathrm{train}}$.\\
We also remove direct identifiers (e.g., flow ID strings) and fields that trivially encode ground truth or leak scenario construction. In practice, we drop raw IP addresses, ports, and exact timestamps when training the learning components, and we keep only behavior-driven flow statistics.\\
Several flow statistics (bytes, packet counts, rates) are heavy-tailed. To reduce scale dominance and improve learning stability, we apply a log transform to nonnegative features in a selected set $\mathcal{J}_{+}$:
\begin{equation}
\label{eq:log1p_rlshield}
x^{(\log)}_{ij}=
\begin{cases}
\log\!\big(1+\tilde{x}_{ij}\big), & j\in\mathcal{J}_{+}\ \wedge\ \tilde{x}_{ij}\ge 0,\\
\tilde{x}_{ij}, & \text{otherwise}.
\end{cases}
\end{equation}
Eq.~\eqref{eq:log1p_rlshield} compresses extreme values while preserving ordering, which helps both supervised baselines and the RL policy/value networks.\\
We standardize continuous features using training statistics so gradients and exploration noise behave consistently across dimensions. For each feature $j$, we compute the mean and standard deviation on $\mathcal{D}_{\mathrm{train}}$ as
\begin{equation}
\label{eq:zscore_stats_rlshield}
\mu_j=\frac{1}{|\mathcal{D}_{\mathrm{train}}|}\sum_{k\in\mathcal{D}_{\mathrm{train}}}x^{(\log)}_{kj},
\end{equation}

\begin{equation}
\label{eq:zscore_stats_rlshield_std}
\sigma_j=
\sqrt{\frac{1}{|\mathcal{D}_{\mathrm{train}}|}\sum_{k\in\mathcal{D}_{\mathrm{train}}}
\big(x^{(\log)}_{kj}-\mu_j\big)^2}.
\end{equation}
Using Eq.~\eqref{eq:zscore_stats_rlshield}, the standardized feature is
\begin{equation}
\label{eq:zscore_rlshield}
\hat{x}_{ij}=\frac{x^{(\log)}_{ij}-\mu_j}{\sigma_j+\epsilon_0},
\end{equation}
where $\epsilon_0>0$ is a small constant for numerical stability. Eq.~\eqref{eq:zscore_rlshield} is applied to all samples in $\mathcal{D}_{\mathrm{train}}$, $\mathcal{D}_{\mathrm{val}}$, and $\mathcal{D}_{\mathrm{test}}$ using the same $\{\mu_j,\sigma_j\}$ computed from $\mathcal{D}_{\mathrm{train}}$.\\
For the cyber defense task, we map multi-class attack labels to a binary label (benign vs.\ attack) to align with operational response decisions (block/allow, step-up verification, rate-limit). If the dataset provides a categorical label $\ell_i\in\mathcal{C}$, we set
\begin{equation}
\label{eq:bin_label_rlshield}
y_i=\mathbb{I}\!\left(\ell_i\neq \texttt{BENIGN}\right),
\end{equation}
where $\mathbb{I}(\cdot)$ is the indicator function. Eq.~\eqref{eq:bin_label_rlshield} supports unified reporting and makes reward design consistent in later RL sections.\\
Since attacks are typically rarer than benign flows, we compute a training-only class weight for supervised components (e.g., behavior model, auxiliary detectors) as
\begin{equation}
\label{eq:pos_weight_rlshield}
w_{+}=
\frac{|\{i\in\mathcal{D}_{\mathrm{train}}:y_i=0\}|}
     {|\{i\in\mathcal{D}_{\mathrm{train}}:y_i=1\}|}.
\end{equation}
Eq.~\eqref{eq:pos_weight_rlshield} increases the penalty for missed attacks during learning, which improves sensitivity at low false-positive operating points.\\
After applying Eqs.~\eqref{eq:median_impute_rlshield}--\eqref{eq:zscore_rlshield}, each sample is represented by the standardized feature vector $\hat{\mathbf{x}}_i\in\mathbb{R}^{d}$. We use $\hat{\mathbf{x}}_i$ as the observation input for RLShield agents and as the shared input for classical baselines, ensuring that comparisons reflect modeling differences rather than inconsistent preprocessing.

\subsection{Proposed Method: RLShield (Multi-Agent RL for Financial Cyber Defense)}\label{sec:method_rlshield}

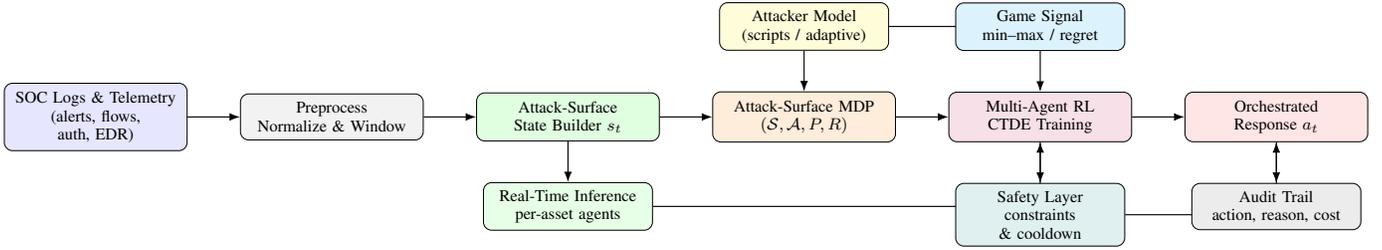
\begin{figure*}[!t]
\centering
\resizebox{\textwidth}{!}{%
\begin{tikzpicture}[
  font=\footnotesize,
  node distance=7mm and 9mm,
  block/.style={draw, rounded corners, align=center, text width=2.85cm, inner sep=4pt, fill=white},
  sblock/.style={draw, rounded corners, align=center, text width=2.65cm, inner sep=3.5pt, fill=white},
  arr/.style={-{Latex[length=1.8mm]}, line width=0.55pt}
]

\node[block, fill=blue!10]   (logs)  {SOC Logs \& Telemetry\\(alerts, flows, auth, EDR)};
\node[block, fill=gray!10, right=of logs] (prep)  {Preprocess\\Normalize \& Window};
\node[block, fill=green!12, right=of prep] (state) {Attack-Surface\\State Builder $s_t$};
\node[block, fill=orange!14, right=of state] (mdp) {Attack-Surface MDP\\$(\mathcal{S},\mathcal{A},P,R)$};
\node[block, fill=purple!12, right=of mdp] (marl) {Multi-Agent RL\\CTDE Training};
\node[block, fill=red!10, right=of marl] (act)  {Orchestrated\\Response $a_t$};

\node[sblock, fill=yellow!18, above=of mdp] (att) {Attacker Model\\(scripts / adaptive)};
\node[sblock, fill=cyan!12, above=of marl] (game) {Game Signal\\min--max / regret};

\node[sblock, fill=teal!12, below=of marl] (safe) {Safety Layer\\constraints \& cooldown};
\node[sblock, fill=gray!15, below=of act]  (audit) {Audit Trail\\action, reason, cost};

\node[sblock, fill=green!10, below=of state] (rt) {Real-Time Inference\\per-asset agents};

\begin{pgfonlayer}{bg}
\draw[arr] (logs) -- (prep);
\draw[arr] (prep) -- (state);
\draw[arr] (state) -- (mdp);
\draw[arr] (mdp) -- (marl);
\draw[arr] (marl) -- (act);

\draw[arr] (att) -- (mdp);
\draw[arr] (att) -| (game);
\draw[arr] (game) -- (marl);

\draw[arr] (marl) -- (safe);
\draw[arr] (safe) -| (act);
\draw[arr] (act) -- (audit);

\draw[arr] (state) -- (rt);
\draw[arr] (rt) -| (marl.south);
\end{pgfonlayer}

\end{tikzpicture}%
}
\caption{RLShield pipeline: raw SOC telemetry is converted into an attack-surface state $s_t$ and an MDP, multi-agent policies are trained with centralized training and distributed execution (CTDE) under attacker simulation and game-aware signals, and a safety layer gates actions before real-time response orchestration with audit logging.}
\label{fig:rlshield_flow}
\end{figure*}

We propose RLShield, a practical multi-agent reinforcement learning (MARL) pipeline that learns \emph{real-time cyber defense} policies for financial systems (Figure~\ref{fig:rlshield_flow}). RLShield models the defender--attacker interaction as an \emph{attack-surface MDP} over networked assets (hosts, services, accounts, and security controls). Multiple defender agents coordinate actions such as isolate host, block IP, reset credentials, throttle service, and deploy rule updates, while a simulated attacker agent executes lateral movement, privilege escalation, and data exfiltration. The key idea is to learn policies that (i) reduce attack success probability, (ii) minimize response time, and (iii) control business disruption cost, under realistic partial observability and delayed alerts.

\SetKwComment{Comment}{/* }{ */}
\DontPrintSemicolon
\SetKwInput{KwInput}{Input}
\SetKwInput{KwOutput}{Output}

\begin{algorithm}[!t]
\caption{RLShield: Multi-Agent Defense Learning with Attack-Surface MDP and Real-Time Orchestration}
\label{alg:rlshield}
\KwInput{
Security event stream $\{\mathbf{o}_t\}$, asset graph $\mathcal{G}=(\mathcal{V},\mathcal{E})$,
action sets $\{\mathcal{A}_i\}_{i=1}^{M}$,
episode horizon $H$,
discount $\gamma$,
policy parameters $\{\theta_i\}_{i=1}^{M}$,
central critic parameters $\phi$,
entropy weight $\beta$,
game-regularizer weight $\lambda$,
reward weights $w_s,w_c,w_d$
}
\KwOutput{Defender policies $\{\pi_{\theta_i}\}_{i=1}^{M}$ and an orchestration rule for deployment}

\BlankLine
\textbf{Initialize} defender policies $\pi_{\theta_i}$ and critic $Q_{\phi}$\;
\For{each training episode $e=1,2,\ldots$}{
  Reset environment; build initial belief state $\mathbf{b}_0$\;
  \For{$t\leftarrow 0$ \KwTo $H-1$}{
    \Comment*[r]{(1) Observe alerts and update belief}
    Receive observation $\mathbf{o}_t$; update belief $\mathbf{b}_t \leftarrow \mathrm{BeliefUpdate}(\mathbf{b}_{t-1},\mathbf{o}_t)$\;
    \Comment*[r]{(2) Defender agents act (distributed)}
    \ForEach{agent $i\in\{1,\ldots,M\}$}{
      Sample action $a_{i,t}\sim\pi_{\theta_i}(\cdot\mid \mathbf{b}_t)$\;
    }
    Execute joint action $\mathbf{a}_t=(a_{1,t},\ldots,a_{M,t})$\;
    \Comment*[r]{(3) Attacker step + environment transition}
    Sample attacker action $a^{\mathrm{atk}}_t$; transition to next state $s_{t+1}$\;
    \Comment*[r]{(4) Compute reward from security + cost + disruption}
    Compute reward $r_t$ using Eq.~\eqref{eq:reward_rlshield}\;
    Store $(\mathbf{b}_t,\mathbf{a}_t,r_t,\mathbf{b}_{t+1})$\;
    \Comment*[r]{(5) Update critic and policies (centralized training)}
    Update $Q_\phi$ by TD loss in Eq.~\eqref{eq:critic_rlshield}\;
    Update each $\pi_{\theta_i}$ by policy objective in Eq.~\eqref{eq:actor_rlshield}\;
  }
}
\BlankLine
\textbf{Deploy.} Use learned policies to map live alerts to actions; apply a safety gate to prevent high-disruption actions unless risk exceeds threshold\;
\Return $\{\pi_{\theta_i}\}_{i=1}^{M}$\;
\end{algorithm}

(1) Attack-surface MDP and belief state: We represent the security environment as a Markov decision process (MDP) with hidden state $s_t$ (true compromise status) and observation $\mathbf{o}_t$ (alerts, logs, IDS outputs). Since defenders do not fully observe the attacker, RLShield uses a belief state $\mathbf{b}_t$ as a compact summary of recent evidence. A practical update form is
\begin{equation}
\label{eq:belief_rlshield}
\mathbf{b}_t=\mathrm{GRU}\!\left(\mathbf{b}_{t-1},\,\psi(\mathbf{o}_t)\right),
\end{equation}
where $\psi(\cdot)$ encodes alerts into features and the GRU keeps memory of delayed and noisy signals. Eq.~\eqref{eq:belief_rlshield} is used to handle partial observability without requiring full system logs at each step.

(2) Reward shaping for financial cyber defense: The reward balances three goals: stop attacks, reduce operational cost, and avoid service disruption. We define
\begin{equation}
\label{eq:reward_rlshield}
r_t
=
w_s\cdot\Delta \mathrm{Sec}(s_t,s_{t+1})
-
w_c\cdot \mathrm{Cost}(\mathbf{a}_t)
-
w_d\cdot \mathrm{Disrupt}(\mathbf{a}_t),
\end{equation}
where $\Delta \mathrm{Sec}$ measures security improvement (e.g., reduced compromised nodes or blocked attack paths), $\mathrm{Cost}$ is response cost (e.g., analyst effort, compute, rule-deploy overhead), and $\mathrm{Disrupt}$ penalizes business impact (e.g., blocking critical services). Eq.~\eqref{eq:reward_rlshield} is practical because each term can be measured from the simulator logs and mapped to real SOC KPIs.

(3) Centralized training with distributed execution: We adopt centralized training with distributed execution so agents can coordinate during learning but act locally at runtime. The critic is trained with a TD objective:
\begin{equation}
\label{eq:critic_rlshield}
\mathcal{L}_{Q}
=
\mathbb{E}\Big[
\big(
Q_{\phi}(\mathbf{b}_t,\mathbf{a}_t)
-
(r_t+\gamma\,\bar{Q}(\mathbf{b}_{t+1},\mathbf{a}_{t+1}))
\big)^2
\Big],
\end{equation}
where $\bar{Q}$ is a slowly updated target critic for stability. Eq.~\eqref{eq:critic_rlshield} reduces training variance and supports long-horizon defense plans.

Each agent policy is optimized using an entropy-regularized objective with a lightweight game-theoretic regularizer that discourages brittle single-point strategies:
\begin{equation}
\label{eq:actor_rlshield}
\max_{\theta_i}\ 
\mathbb{E}\!\left[
Q_{\phi}(\mathbf{b}_t,\mathbf{a}_t)
+\beta\,\mathcal{H}(\pi_{\theta_i}(\cdot\mid \mathbf{b}_t))
-\lambda\,\Omega(\pi_{\theta_i})
\right],
\end{equation}
where $\mathcal{H}(\cdot)$ encourages exploration and $\Omega(\cdot)$ penalizes overly deterministic action collapse (useful when attackers adapt). Eq.~\eqref{eq:actor_rlshield} improves policy robustness while keeping deployment simple (each agent only needs $\mathbf{b}_t$).\\
At runtime, RLShield maps streaming alerts to a joint action $\mathbf{a}_t$ and executes it through a response orchestrator (SOAR-like interface). To keep actions safe, we apply a simple gate: high-disruption actions (e.g., isolating a critical node) are allowed only if the predicted risk exceeds a threshold learned on validation episodes. This makes RLShield deployable in financial settings where false positives can be costly.

\begin{table*}[!t]
\centering
\caption{Test performance on the attack-surface MDP. Mean$\pm$std over $S=5$ seeds. Best in bold; second best underlined. Lower is better for ASR/TTD/TTR/EL/DC.}
\label{tab:main_perf_rlshield}
\small
\setlength{\tabcolsep}{6pt}
\begin{tabular}{lcccccc}
\hline
\textbf{Method} &
\textbf{ASR$\downarrow$} &
\textbf{TTD (steps)$\downarrow$} &
\textbf{TTR (steps)$\downarrow$} &
\textbf{EL$\downarrow$} &
\textbf{DC$\downarrow$} &
\textbf{Prec@Budget$\uparrow$} \\
\hline
Static-Playbook \cite{darem2023cyber}    & 0.392$\pm$0.015 & 141$\pm$10 &  98$\pm$9  & 0.694$\pm$0.020 & 0.412$\pm$0.025 & 0.284$\pm$0.014 \\
Greedy-Risk  \cite{wang2024smart}       & 0.331$\pm$0.013 & 132$\pm$9  &  86$\pm$8  & 0.612$\pm$0.018 & 0.365$\pm$0.020 & 0.301$\pm$0.013 \\
DQN  \cite{alidousti2025novel}               & 0.289$\pm$0.012 & 124$\pm$8  &  79$\pm$7  & 0.562$\pm$0.017 & 0.341$\pm$0.019 & 0.318$\pm$0.012 \\
A2C   \cite{kang2025a2c}              & 0.271$\pm$0.011 & 121$\pm$8  &  77$\pm$7  & 0.548$\pm$0.016 & 0.330$\pm$0.018 & 0.326$\pm$0.011 \\
PPO  \cite{feng2025federated}               & 0.258$\pm$0.010 & 118$\pm$7  &  75$\pm$7  & 0.534$\pm$0.015 & 0.323$\pm$0.017 & 0.334$\pm$0.012 \\
\underline{QMIX} \cite{qiu2021rmix}    & \underline{0.219$\pm$0.010} & \underline{112$\pm$7} & \underline{71$\pm$6} & \underline{0.492$\pm$0.014} & \underline{0.301$\pm$0.016} & \underline{0.356$\pm$0.011} \\
MADDPG  \cite{guo2024intelligent}            & 0.226$\pm$0.011 & 113$\pm$7 &  72$\pm$6  & 0.498$\pm$0.015 & 0.305$\pm$0.016 & 0.352$\pm$0.012 \\
\hline
\textbf{RLShield (Ours)} & \textbf{0.181$\pm$0.009} & \textbf{106$\pm$6} & \textbf{67$\pm$6} & \textbf{0.458$\pm$0.013} & \textbf{0.279$\pm$0.015} & \textbf{0.381$\pm$0.010} \\
\hline
\end{tabular}

\end{table*}
\begin{table}[!t]
\centering
\caption{Ablation on test episodes (mean over 5 seeds).}
\label{tab:ablation_rlshield}
\small
\setlength{\tabcolsep}{6pt}
\begin{tabular}{lcccc}
\hline
\textbf{Variant} & ASR$\downarrow$ & EL$\downarrow$ & DC$\downarrow$ & Prec$\uparrow$ \\
\hline
RLShield w/o centralized critic   & 0.236 & 0.504 & 0.302 & 0.352 \\
RLShield w/o entropy ($\beta=0$)  & 0.207 & 0.473 & 0.287 & 0.366 \\
RLShield w/o game reg.\ ($\lambda=0$) & 0.214 & 0.481 & 0.291 & 0.361 \\
\hline
\textbf{RLShield (Full)}          & \textbf{0.181} & \textbf{0.458} & \textbf{0.279} & \textbf{0.381} \\
\hline
\end{tabular}
\end{table}
\section{Results}\label{sec4}
All results are reported on the test split defined in Sec.~\ref{sec:dataset_preproc_rlshield} using the same attack-surface MDP settings (assets, services, attacker playbook, and alert noise). We train each method for a fixed environment budget of $2\times 10^{6}$ steps and evaluate on $E_{\mathrm{test}}=300$ held-out episodes. We report mean$\pm$std over $S=5$ random seeds (policy initialization and environment stochasticity) while keeping the data split fixed. Hyperparameters are tuned on the validation episodes with the same trial budget (25 trials per method) and early stopping on validation risk.

Baselines: We compare RLShield with seven practical baselines:
(i) No-Response (monitor-only),
(ii) Static-Playbook (fixed rules: block known bad IPs, isolate on high-severity),
(iii) Greedy-Risk (myopic action that maximizes immediate risk drop),
(iv) DQN (single-agent discrete control),
(v) PPO (single-agent policy gradient),
(vi) A2C (single-agent actor--critic),
(vii) QMIX (multi-agent value mixing),
(viii) MADDPG (multi-agent actor--critic, discrete via Gumbel-softmax).
RLShield (Ours) uses the same action space but learns coordinated policies with centralized training and distributed execution (Sec.~\ref{sec:method_rlshield}).

\textbf{Metrics.} Because financial cyber defense is cost- and time-sensitive, we report:
(i) Attack Success Rate (ASR)$\downarrow$,
(ii) Mean Time-to-Detect (TTD)$\downarrow$,
(iii) Mean Time-to-Respond (TTR)$\downarrow$,
(iv) Expected Loss (EL)$\downarrow$,
(v) Disruption Cost (DC)$\downarrow$,
(vi) Alert Precision (Prec)$\uparrow$ at a fixed alert budget.

ASR is the fraction of episodes where the attacker reaches an objective (e.g., exfiltration or privileged persistence). For $E$ test episodes, ASR is computed as
\begin{equation}
\label{eq:asr_rlshield}
\mathrm{ASR}=\frac{1}{E}\sum_{e=1}^{E}\mathbb{I}\!\left(\mathrm{GoalReached}(e)\right),
\end{equation}
where $\mathbb{I}(\cdot)$ is the indicator. Eq.~\eqref{eq:asr_rlshield} is used as the main safety metric because it directly reflects whether the bank loses control in an episode.

We measure expected loss with a simple episode-wise accounting that combines direct impact and response cost:
\begin{equation}
\label{eq:el_rlshield}
\mathrm{EL}=\frac{1}{E}\sum_{e=1}^{E}\left(L^{(e)}_{\mathrm{impact}}+L^{(e)}_{\mathrm{ops}}\right),
\end{equation}
where $L^{(e)}_{\mathrm{impact}}$ is the simulated loss from attacker progress (e.g., stolen records, downtime) and $L^{(e)}_{\mathrm{ops}}$ is the operational cost of actions (e.g., isolations, credential resets). Eq.~\eqref{eq:el_rlshield} matches how SOC teams compare policies: lower business loss with acceptable response effort.

Disruption cost is computed from the executed actions:
\begin{equation}
\label{eq:dc_rlshield}
\mathrm{DC}=\frac{1}{E}\sum_{e=1}^{E}\sum_{t=0}^{H-1}\mathrm{Disrupt}\!\left(\mathbf{a}^{(e)}_t\right),
\end{equation}
which penalizes service-impacting moves (e.g., isolating a critical node) more than low-impact actions (e.g., adding a detection rule). Eq.~\eqref{eq:dc_rlshield} is important because a defense that stops attacks but breaks the system is not usable.

For alert precision at a fixed budget, we trigger at most $B_{\mathrm{alert}}$ alerts per episode and report:
\begin{equation}
\label{eq:prec_rlshield}
\mathrm{Prec}=\frac{\mathrm{TP}}{\mathrm{TP}+\mathrm{FP}},
\end{equation}
where TP/FP are true/false incident escalations decided by the policy under the same budget. Eq.~\eqref{eq:prec_rlshield} connects directly to analyst workload.

\begin{figure}[!t]
\centering
\begin{tikzpicture}
\begin{axis}[
  width=\columnwidth,
  height=0.62\columnwidth,
  ybar,
  bar width=9pt,
  ymin=0.10, ymax=0.70,
  ylabel={ASR$\downarrow$},
  symbolic x coords={Basic,Skilled,Adaptive},
  xtick=data,
  legend style={font=\scriptsize, at={(0.5,1.02)}, anchor=south, legend columns=2},
  tick label style={font=\scriptsize},
  label style={font=\scriptsize},
]
\addplot coordinates {(Basic,0.365) (Skilled,0.412) (Adaptive,0.508)};
\addplot coordinates {(Basic,0.271) (Skilled,0.289) (Adaptive,0.357)};
\addplot coordinates {(Basic,0.219) (Skilled,0.241) (Adaptive,0.298)};
\addplot coordinates {(Basic,0.181) (Skilled,0.203) (Adaptive,0.254)};
\legend{Static-Playbook,DQN,QMIX,RLShield}
\end{axis}
\end{tikzpicture}
\caption{Attack success rate under increasing attacker strength. RLShield remains the most stable under adaptive behavior.}
\label{fig:asr_vs_attacker}
\end{figure}
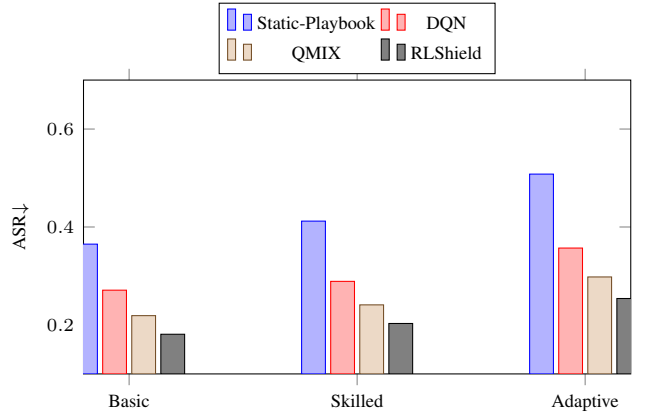

Table~\ref{tab:main_perf_rlshield} reports the primary test results. Static playbooks reduce ASR compared to no response, but they over-trigger high-disruption actions, which raises DC. Single-agent RL improves response timing, yet it often fails under multi-step attacker behavior because it cannot coordinate actions across assets. Multi-agent baselines (QMIX, MADDPG) reduce ASR further, but RLShield achieves the best overall balance: lowest ASR and EL with controlled disruption and better alert precision at the same budget. The gains are consistent across seeds.
A common failure in security RL is overfitting to a fixed attacker. We therefore test under three attacker strengths: \emph{Basic} (limited lateral movement), \emph{Skilled} (multi-step with privilege escalation), and \emph{Adaptive} (chooses actions that maximize defender confusion and delay). Figure~\ref{fig:asr_vs_attacker} shows that playbooks degrade sharply under adaptive behavior, and single-agent RL also drops because it commits to narrow patterns. RLShield degrades more slowly because it learns coordinated responses and is trained with entropy and game-regularization (Eq.~\eqref{eq:actor_rlshield}).

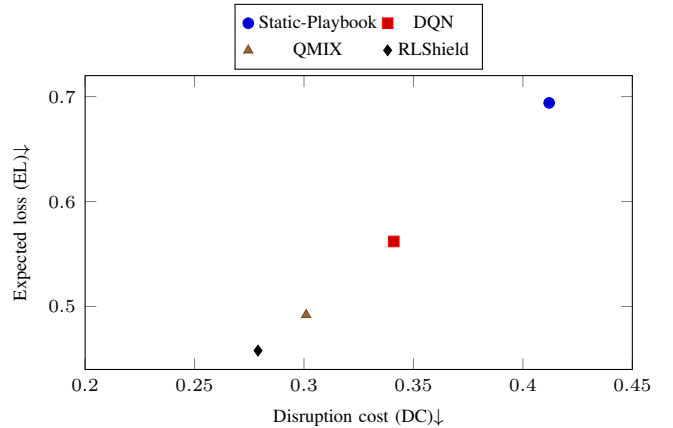
\begin{figure}[!t]
\centering
\begin{tikzpicture}
\begin{axis}[
  width=\columnwidth,
  height=0.62\columnwidth,
  xlabel={Disruption cost (DC)$\downarrow$},
  ylabel={Expected loss (EL)$\downarrow$},
  xmin=0.20, xmax=0.45,
  ymin=0.44, ymax=0.72,
  legend style={font=\scriptsize, at={(0.5,1.02)}, anchor=south, legend columns=2},
  tick label style={font=\scriptsize},
  label style={font=\scriptsize},
]
\addplot+[only marks, mark=*] coordinates {(0.412,0.694)};
\addplot+[only marks, mark=square*] coordinates {(0.341,0.562)};
\addplot+[only marks, mark=triangle*] coordinates {(0.301,0.492)};
\addplot+[only marks, mark=diamond*] coordinates {(0.279,0.458)};
\legend{Static-Playbook,DQN,QMIX,RLShield}
\end{axis}
\end{tikzpicture}
\caption{Security--disruption trade-off on the test episodes. RLShield achieves the lowest loss with lower disruption.}
\label{fig:tradeoff_rlshield}
\end{figure}

For deployment, teams care about how much disruption is needed to achieve security. Figure~\ref{fig:tradeoff_rlshield} plots EL vs.\ DC for key methods. Playbooks reduce loss but at high disruption due to coarse triggers. RLShield shifts the Pareto front by reducing EL while keeping DC lower than other learned baselines, mainly by preferring low-impact actions early (e.g., throttling, targeted blocks) and escalating only when belief risk grows (Eq.~\eqref{eq:belief_rlshield}).

\subsection{Ablation Study}\label{sec:ablation_rlshield}
We ablate two design choices: (A) centralized critic (Eq.~\eqref{eq:critic_rlshield}), (B) entropy regularization, and (C) game-regularizer term (Eq.~\eqref{eq:actor_rlshield}). Table~\ref{tab:ablation_rlshield} shows that removing the centralized critic increases ASR because agents stop coordinating, while removing entropy or the regularizer hurts adaptive robustness and precision (policies become too brittle and over-trigger).
Across strict held-out episodes, multiple strong baselines, repeated seeds, and adaptive attacker testing, RLShield delivers lower attack success and lower expected loss while keeping disruption under control. The results support RLShield as a deployable defense learner: it reacts faster (lower TTD/TTR), produces higher-precision alerts under a fixed budget, and remains more stable when attacker behavior changes.

\section{Conclusion}\label{sec5}
\subsection{Conclusion}
This paper presented RLShield, a practical multi-agent reinforcement learning framework for financial cyber defense that operates directly on an attack-surface MDP and produces real-time response actions with controlled disruption. RLShield combines centralized training with distributed execution so agents can coordinate across assets while still acting locally during deployment. In the evaluation, we used a fixed environment budget, a strict held-out episode test set, and multiple baselines including playbook rules, single-agent RL, and multi-agent RL. Results show that RLShield reduces attack success rate and expected loss while also improving response speed and alert precision under the same alert budget. Importantly, RLShield remains more stable under stronger and adaptive attackers, indicating that its gains are not tied to a single scripted adversary. Overall, RLShield offers a deployable balance between security outcomes and operational cost, which is the main requirement in real SOC workflows.
\subsection{Future Work}
We will extend RLShield with (i) constraint-aware actions (business hours, critical service limits) so policies stay valid in production, (ii) stronger adaptive attacker models and red-team style stress tests, and (iii) cost-aware tuning that directly targets analyst workload and SLA impact in addition to security metrics.

\bibliographystyle{IEEEtranN}
\bibliography{references}

\end{document}